\documentclass[10pt]{iopart}

\usepackage{graphicx} 
\usepackage[usenames,dvipsnames]{color} 
\expandafter\let\csname equation*\endcsname\relax
\expandafter\let\csname endequation*\endcsname\relax
\usepackage{amsmath} 
\usepackage[urlcolor=blue, hyperindex, colorlinks, bookmarks=true,linkcolor=black,citecolor=black]{hyperref} 
\usepackage{dcolumn}
\usepackage{amssymb} 
\usepackage{soul} 
\usepackage{ifthen} 
\usepackage{bbm}
\usepackage{comment}
\usepackage[ampersand]{easylist}
\usepackage{todonotes}
\usepackage{upgreek}

\def\l{\left}
\def\r{\right}

\def\be#1\ee{\begin{equation}#1\end{equation}}
\def\ba#1\ea{\begin{align}#1\end{align}}
\def\bg#1\eg{\begin{gather}#1\end{gather}}
\def\t{\text}

\newcommand{\mb}[1]{\mathbf{#1}}

\newcommand{\uwiqc}{Institute for Quantum Computing, and Department of Physics and Astronomy,
University of Waterloo, Waterloo, ON, Canada N2L 3G1}

\newcommand{\uwwin}{Waterloo Institute for Nanotechnology, University of Waterloo, Waterloo, ON, Canada N2L 3G1}
\def\shownoteal{1} 
\newcommand{\nal}[1]{\ifthenelse{\shownoteal=1}{\textcolor{red}{[[#1]]}}{}}
\def\shownoteay{1} 
\newcommand{\nay}[1]{\ifthenelse{\shownoteay=1}{\textcolor{orange}{[[#1]]}}{}}
\def\showaddmat{0} 
\newcommand{\addmat}[1]{\ifthenelse{\showaddmat=1}{\textcolor{Gray}{[[#1]]}}{}}

\begin{document}

\title[Optimizing for periodicity to calibrate flux crosstalk]{Optimizing for periodicity: a model-independent approach to flux crosstalk calibration for superconducting circuits}

\author{
X. Dai$^1$, 
R. Trappen$^1$, 
R. Yang$^1$, 
S. M. Disseler$^2$,
J. I. Basham$^2$,
J. Gibson$^3$, 
A. J. Melville$^2$, 
B. M. Niedzielski$^2$, 
R. Das$^2$, 
D. K. Kim$^2$,
J. L. Yoder$^2$, 
S. J. Weber$^2$, 
C. F. Hirjibehedin$^2$,
D. A. Lidar$^4$,
and A. Lupascu$^{1,5}$
}
\address{$^1$\uwiqc}
\address{$^2$Lincoln Laboratory, Massachusetts Institute of Technology, Lexington, Massachusetts, 02421, USA}
\address{$^3$Department of Physics and Astronomy, Dartmouth College, Hanover, NH 03755, USA}
\address{$^4$Departments of Electrical \& Computer Engineering, Chemistry, and Physics, and Center for Quantum Information Science \& Technology, University of Southern California, Los Angeles, California 90089, USA}
\address{$^5$\uwwin}
\ead{x35dai@uwaterloo.ca; alupascu@uwaterloo.ca}

\date{ \today}

\begin{abstract}
Flux tunability is an important engineering resource for superconducting circuits. Large-scale quantum computers based on flux-tunable superconducting circuits face the problem of flux crosstalk, which needs to be accurately calibrated to realize high-fidelity quantum operations. Typical calibration methods either assume that circuit elements can be effectively decoupled and simple models can be applied, or require a large amount of data. Such methods become ineffective as the system size increases and circuit interactions become stronger. Here we propose a new method for calibrating flux crosstalk, which is independent of the underlying circuit model. Using the fundamental property that superconducting circuits respond periodically to external fluxes, crosstalk calibration of $N$ flux channels can be treated as $N$ independent optimization problems, with the objective functions being the periodicity of a measured signal depending on the compensation parameters. We demonstrate this method on a small-scale quantum annealing circuit based on superconducting flux qubits, achieving comparable accuracy with previous methods. We also show that the objective function usually has a nearly convex landscape, allowing efficient optimization. 
\end{abstract}
\maketitle

\section{Introduction}
Superconducting circuits have become one of the most promising platforms for realization of large-scale quantum computers~\cite{arute_2019_quantumsupremacyusing,kjaergaard_2020_superconductingqubitscurrent,wu_2021_strongquantumcomputationala}. One of the main advantages of superconducting circuits is their versatile design space, allowing fast, high-fidelity operation and readout while suppressing noise sensitivity~\cite{mooij_1999_josephsonpersistentcurrentqubit,koch_2007_chargeinsensitivequbitdesign,manucharyan_2009_fluxoniumsinglecooperpair,fei_2018_tunablecouplingscheme,gyenis_2021_movingtransmonnoiseprotected,ye_2021_engineeringpurelynonlinear}. A key engineering resource for superconducting circuits is in-situ tunability by external flux biasing, which has been used to realize high-fidelity gates~\cite{chen_2014_qubitarchitecturehigh,reagor_2018_demonstrationuniversalparametric,googleaiquantum_2020_demonstratingcontinuousset,xu_2020_highfidelityhighscalabilitytwoqubit,sung_2021_realizationhighfidelitycz,stehlik_2021_tunablecouplingarchitecture,negirneac_2021_highfidelitycontrolledgate,zhao_2022_quantumcrosstalkanalysis}, tunable couplers~\cite{vanderploeg_2007_controllablecouplingsuperconducting,harris_2007_signmagnitudetunablecoupler,allman_2010_rfsquidmediatedcoherenttunablea,srinivasan_2011_tunablecouplingcircuit,baust_2015_tunableswitchablecoupling,weber_2017_coherentcoupledqubits,menke_2022_demonstrationtunablethreebody}, avoiding frequency crowding~\cite{arute_2019_quantumsupremacyusing} and two-level-system defects~\cite{klimovFluctuationsEnergyRelaxationTimes2018}, as well as to build programmable quantum annealers~\cite{johnson_2011_quantumannealingmanufactured}. 

Operating large-scale superconducting-circuits based quantum computers requires accurate calibration of the system parameters and control crosstalk~\cite{neill_2021_accuratelycomputingelectronic,king_coherent_2022}. For flux control specifically, crosstalk arises due to the physical proximity between circuit elements and control lines, as well as reasons associated with the electromagnetic environment hosting the circuits, such as ground loops. For most large-scale superconducting circuits today, which are based on transmons, solving the calibration problem is often helped by the fact that transmons interact via the charge degree of freedom and the interaction strength is weak~\cite{kounalakis_2018_tuneablehoppingnonlinear,neill_2018_blueprintdemonstratingquantum,abrams_2019_methodsmeasuringmagnetic,krinner_2021_realizingrepeatedquantum}. For annealers based on flux qubits, calibration is more challenging because of the strong flux interaction between circuit elements, which makes it hard to directly measure the coupling between bias lines and flux loops~\cite{daiCalibrationFluxCrosstalk2021,tennant_demonstration_2022,menke_2022_demonstrationtunablethreebody}. 

In this work we introduce a new approach to calibrate crosstalk between flux biases for superconducting circuits. Relying on the fundamental property that superconducting circuits respond periodically to external flux biases~\cite{london_1961_superfluidsvolumemacroscopic,byers_1961_theoreticalconsiderationsconcerning}, orthogonal control of flux biases can be obtained by optimizing for the periodic response of the circuit relative to each target bias coordinate, over the compensation parameters from all other bias lines. The periodicity analysis can be automated, allowing a closed loop optimization to be performed. Unlike conventional calibration methods which require either a simple model to describe the measurement signal, or high resolution scans, the periodicity optimization approach is circuit-model independent, and does not require high resolution scans. We demonstrate this approach on a small quantum annealing circuit, achieving a comparable accuracy with previous calibration methods on a subcircuit with three loops~\cite{daiCalibrationFluxCrosstalk2021}. 

The paper is organized as follows. In Section~\ref{sec:CrosstalkProblem} we introduce the notations used to describe the flux crosstalk problem and state the periodicity condition. In Section~\ref{sec:Traditional} we briefly review the principles behind earlier crosstalk calibration approach. In Section~\ref{sec:Method} we discuss the framework for crosstalk calibration based on maximizing periodicity, as well as the metric used to quantify periodicity. In Section~\ref{sec:Implementation} we discuss the experimental results of implementing this method on a quantum annealing circuit, followed by conclusions in Section~\ref{sec:Conclusion}.

\section{Flux crosstalk and periodicity}\label{sec:CrosstalkProblem}
The properties of superconducting circuits depend on the external flux biases of the superconducting loops. For a superconducting circuits with $N$ flux bias loops, the external fluxes are usually controlled by $N$ bias lines, which are mutually coupled to the flux loops. We denote the external flux bias in loop $i$, reduced by the flux quantum $\Phi_0$, as $f_i$, and the corresponding bias line current as $I_i$. The fluxes $\{f_i\}$ in all loops and currents $\{I_i\}$ on all bias lines can be written as vectors $\mb{f}$ and $\mb{I}$ respectively, and they are in general related by a linear transformation
\begin{align}
\mb{f} = \mb{M}\mb{I} + \mb{f}_0,
\end{align}
where $\mb{M}$ is the $N\times N$ mutual matrix describing the coupling between bias lines and flux loops, and $\mb{f}_0$ is the vector of flux offsets arising from spurious sources. Often, and in particular in the context of our experiment, bias currents are controlled by voltage sources. For a more direct representation of the experiment, we will refer to the relation between fluxes and voltages, written as
\begin{align}
    \mb{f}=\mb{C}\mb{V}+\mb{f}_0,
\end{align}
where $\mb{V}$ is the vector of voltages with each element controlling the corresponding element in $\mb{I}$, and $\mb{C}=\mb{M}\mb{R}^{-1}$ with $\mb{R}$ a diagonal matrix consisting of the resistances between the voltage sources and the bias lines. From here onward we will work with voltage controls and the crosstalk matrix $\mb{C}$. 

Measurements on the superconducting qubits can be considered as a function mapping the flux biases to the signal $R_l$, where $l$ denotes a particular readout channel. Note that the number of readout channel is not restricted to the number of physical signal processing units; rather each channel corresponds to reading out the signal of an experiment, with a particular set of experimental parameters. The experiment could consist of one quadrature of a transmission measurement at a particular frequency or more complex experiments involving microwave excitations of the system. 

Superconducting circuits respond periodically to external bias fluxes, with the period of one flux quantum. We denote the readout data as $\mb{R}$, which is a vector with dimension of the number of readout channels. The periodicity condition can be formally stated as
\begin{align}
    \mb{R}(\{f_k\})=\mb{R}(\{f_k+m_k\}), \forall k=1,2,\dots,N
\end{align}
where $m_k$ is an integer.

\section{Translation-based approach to flux crosstalk calibration}\label{sec:Traditional}
Most previously developed approaches to flux crosstalk calibration assume that one can identify a particular readout channel $l$ that depends on the external flux in a single loop $i$, $R_l(\{f_k\})\approx R_l(f_i)$. This allows estimating the coupling coefficient from bias line $j$ to loop $i$, $C_{ij}$, by measuring the translation of $R_l$ as a function of $V_j$. For this reason we denote such calibration method as the translation-based approach. When a simple model for $R_l(f_i)$ exists, the method becomes particularly effective as one only requires measurements at a few voltage bias values to extract the coupling parameters, and the model can be fitted to the data to obtain the coupling $C_{ij}$. This is the case for many calibration methods used in tunable transmons, with the readout channel being the frequency of the transmon or its readout resonator~\cite{kounalakis_2018_tuneablehoppingnonlinear,abrams_2019_methodsmeasuringmagnetic}, or the Ramsey phase shift~\cite{neill_2018_blueprintdemonstratingquantum,abrams_2019_methodsmeasuringmagnetic,krinner_2021_realizingrepeatedquantum}. However, this method would only work if the circuit elements interact weakly, and each superconducting loop can be sufficiently decoupled from the other loops. We also note that the work presented in Ref.~\cite{braumuller_2022_probingquantuminformation} uses an optimization-based crosstalk calibration approach, however, this too relies on simplifying the full superconducting circuit to an effective description in terms of weakly coupled bosonic modes.

Recently, an iterative version of the translation-based approach was developed to tackle the issue of strong circuit interactions~\cite{daiCalibrationFluxCrosstalk2021}. The idea is that in the first iteration of the flux crosstalk calibration, one can assume the readout signal depends on only one flux bias, which changes due to the coupling from the bias lines alone and not due to inductive coupling to other loops. In each new iteration, the control coordinates to be swept become the estimated flux coordinates from the previous iteration, which allows one to gradually decouple the different superconducting loops and converge towards the true crosstalk. However, due to the strong circuit interactions, which are hard to model, this method often requires both high resolution scans and a large number of iterations to calibrate the crosstalk accurately.

\section{Periodicity maximization approach}\label{sec:Method}
In this section we introduce the periodicity optimization approach to crosstalk calibration. We will first discuss the framework used to treat the calibration task as an optimization problem. Then we will discuss the measurement and analysis required to quantify periodicity.

\subsection{Crosstalk calibration as an optimization problem}
The task of crosstalk calibration is to obtain estimates of the coupling matrix $\mb{C}$ and independent control of the external flux biases. This is equivalent to finding $N$ independent control coordinates, such that the circuit responds periodically to changes in each of them. To do this, we break the calibration task into $N$ independent optimization problems, as described below. 

We start by introducing initial estimates of the crosstalk and flux offsets, given by $\mb{C}^\t{init}$ and $\mb{f}_0^{\t{init}}$. Introducing them makes it convenient to discuss the optimization with or without prior knowledge on the same footing. When no prior knowledge is available, the initial estimates are identity and zeros for the crosstalk matrix and flux offsets respectively. The initial estimates allow us to define the initial control coordinates $\mb{f}^{\t{init}}$,
\begin{align}
    \mb{f}^{\t{init}} = \mb{C}^{\t{init}}\mb{V}+\mb{f}_0^{\t{init}}.\label{eq:finit}
\end{align}
The initial control coordinates $\mb{f}^{\t{init}}$ are related to the actual fluxes $\mb{f}$ via the residual crosstalk and flux offsets,
\begin{align}
    \mb{f} = \mb{C}^{\t{res}}\mb{f}^\t{init}+\mb{f}_0^{\t{res}},\label{eq:f}
\end{align}
where $\mb{C}^\t{res}=\mb{C}(\mb{C}^\t{init})^{-1}, \mb{f}_0^\t{res}=\mb{f}_0-\mb{C}^\t{res}\mb{f}_0^\t{init}$.

\begin{figure}
    \centering
    \includegraphics[width=4.5in]{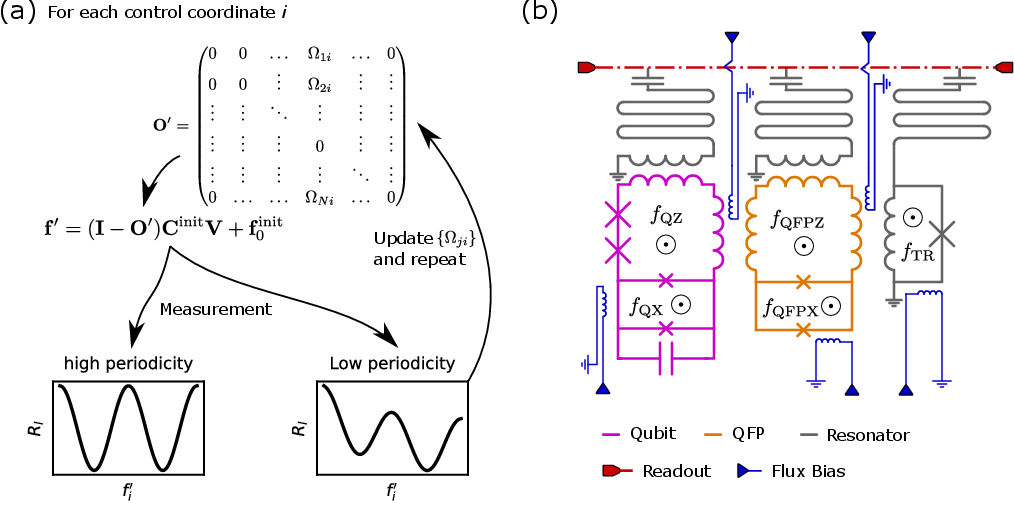}
    \caption{(Color Online) (a)Schematic representation of the optimization step. For each loop $i$, the optimization parameters are elements of a trial compensation matrix $\mb{O}^\prime$, which defines the trial flux coordinates $\mb{f}^\prime$. Then the  measurement is done by sweeping $f_i^\prime$ and the periodicity of the measurement signal is determined. If the periodicity is high, the compensation parameters give good estimates of the ratio between the crosstalk matrix elements, otherwise, the compensation parameters are updated and the optimization is repeated until periodicity is high. (b) Schematic of the subcircuit of the device measured, with the tunable flux qubit on the left (purple), the quantum flux parametron (QFP) in the middle (yellow), and the tunable resonator on the right (grey). In addition, the qubit and the QFP are each coupled to a fixed-frequency resonator (grey). All resonators are coupled to a joint feedline (red). External flux biases in the loops are controlled via the on-chip bias lines (blue). }
    \label{fig:DeviceFlowChart}
\end{figure}

To calibrate the $i$th control coordinate, we define a trial control flux variable $\mb{f}^\prime$, 
\begin{align}
\mb{f}^\prime=(\mb{I}-\mb{O}^\prime)\mb{C}^\t{init}\mb{V}+\mb{f}_0^{\t{init}},\label{eq:fprime}
\end{align}
where the compensation matrix $\mb{O}^\prime$ has elements $O_{jk}^\prime$ with
\begin{align}
    O^\prime_{jk}=
    \begin{cases}
    0, &\t{if}~k\neq i~\t{or}~j= k,\\
    \Omega_{jk}, &\t{otherwise}.
    \end{cases}
\end{align}
There are $N-1$ non-zero elements in the matrix $\mb{O}^\prime$, denoted as $\{\Omega_{ji}\}$. These are the compensation parameters to be optimized when calibrating the $i$th control coordinate. The objective of the optimization problem is to maximize the periodicity of the measured signal when varying the $i$th coordinate of the trial flux, $f^\prime_i$. This can be done by performing measurements varying $f^\prime_i$, and quantifying the periodicity using the metric discussed in Sec.~\ref{sec:Periodicity}. A schematic for one iteration of the optimization step is shown in figure~\ref{fig:DeviceFlowChart}(a).

The maximum periodicity of the signal is achieved when compensation parameters satisfy specific relations relative to the residual crosstalk $\mb{C}^\t{res}$. To see this, consider the relation between the trial control fluxes and the actual fluxes, which follows from Eq.~(\ref{eq:finit},~\ref{eq:f},~\ref{eq:fprime}),
\begin{align}
    \mb{f} = \mb{C}^{\t{res}}(\mb{I}+\mb{O}^\prime)\mb{f}^\prime + \mb{f}_0^\prime,
\end{align}
where $\mb{f}_0^\prime=\mb{f}_0-\mb{C}^\t{res}(\mb{I}+\mb{O}^\prime)\mb{f}_0^\t{init}$. It can be seen that when the following condition is satisfied
\begin{align}
 \Omega_{ji}=\frac{[(\mb{C}^\t{res})^{-1}]_{ji}}{[(\mb{C}^\t{res})^{-1}]_{ii}},~\forall j\neq i,\label{eqn:OptimumCompensation}
\end{align}
one has 
\begin{align}
    f_i&=\frac{f^\prime_i}{[(\mb{C}^{\t{res}})^{-1}]_{ii}}+\sum_{j\neq i}C^\text{res}_{ij}f_j^\prime+f^\prime_{0,i}\label{eq:Optimumfi}\\
    f_{l\neq i}&=\sum_{j\neq i}C^\text{res}_{lj}f_j^\prime+f^\prime_{0,l},\label{eq:Optimumfl}
\end{align}
where $f_i,f^\prime_i,f^\prime_{0,i}$ and $C^\text{res}_{ij}$ are elements of $\mb{f},\mb{f}^\prime,\mb{f}^\prime_0$ and $\mb{C}^\text{res}$ respectively. The relations between the actual fluxes $\mb{f}$ and trial control fluxes $\mb{f}^\prime$ given by Eqs.~(\ref{eq:Optimumfi},~\ref{eq:Optimumfl}) indicate that when the $i$th control flux $f_i^\prime$ is being varied, only the $i$th actual flux $f_i$ changes. In other words, the residual crosstalk from the $i$th control coordinate to other coordinates $l\neq i$ is completely cancelled out by setting the compensation parameters $\{\Omega_{ji}\}$ satisfying Eq.~\ref{eqn:OptimumCompensation}. Since the circuit response is periodic to each flux $f_i$ with period $1$, the circuit also responds periodically with respect to $f^\prime_i$, with period $[(\mb{C}^{\t{res}})^{-1}]_{ii}$. Hence, optimizing the periodicity for the $i$th coordinate gives the optimized compensation parameters approximately satisfying Eq.~\ref{eqn:OptimumCompensation}, and they are denoted as $\Omega_{ji}^\prime$. After completing the optimization for all $N$ control coordinates, we obtain $N(N-1)$ optimized compensation parameters, and another $N$ parameters corresponding to the periods of the $N$ control coordinates. Using Eq.~\ref{eqn:OptimumCompensation}, estimates for residual crosstalk matrix can be obtained, which we denote as $\mb{C}^{\t{res}\prime}$.

\subsection{Quantifying periodicity}\label{sec:Periodicity}
The objective function for the optimization is the periodicity of the readout data with respect to $f^\prime_i$. To measure the periodicity, the readout data is collected sweeping a large enough range of $f^\prime_i$ to cover a few periods, while keeping $\{f^\prime_{j\neq i}\}$ fixed. The readout data is denoted as $R_l(f^\prime_{i,s})$, where $s=1,2,\dots ,m$ goes over the values of $f_i^\prime$ taken during the sweep and $m$ is the total number of $f_i^\prime$ steps. Readout data from different channels is first normalized, by applying the operation
\begin{align}
    R_l(f^\prime_{i,s})\xrightarrow[]{}\frac{R_l(f^\prime_{i,s})-\overline{R_l}}{\sqrt{\sum_s\left[R_l(f^\prime_{i,s})-\overline{R_l}\right]^2}},
\end{align}
where $\overline{R_l}$ is the average of the readout data from channel $l$ over all values of $f_i^\prime$ taken. The periodicity can be quantified by first computing the correlation of the signal and its own with a translation of $t$ steps along the $f_i^\prime$ coordinates. Defining the translated signal as 
\begin{align}
    R_{l,t}(f^\prime_{i,s})=R_l(f^\prime_{i,s+t}),
\end{align}
the correlation is
\begin{align}
    \rho_i(t\delta) &= \frac{\sum_{l,s\in\mathcal{S}} \l[R_l(f^\prime_{i,s})-\overline{R_l}\r]\l[R_{l,t}(f^\prime_{i,s})-\overline{R_{l,t}}\r]}{\sqrt{\sum_{l,s\in\mathcal{S}}\l[R_l(f^\prime_{i,s})-\overline{R_l}\r]^2\sum_{l,s\in\mathcal{S}}\l[R_{l,t}(f^\prime_{i,s})-\overline{R_{l,t}}\r]^2}}~\text{and}\label{eq:Correlation}\\
    \mathcal{S}&=\{1,2,\dots,m-t\},
\end{align}
where $\delta$ is the step size of the $f_i^\prime$ sweep and $t$ is an integer for the translation considered. The $\overline{R_{l}},\overline{R_{l,t}}$ refer to averages of the readout data over $\mathcal{S}$ for a particular readout channel $l$. From the definition of correlation, we have the range of $\rho\in[-1,1]$, with $1$ for perfect correlation, $-1$ for perfect anti-correlation, and $0$ for no correlation. 

The correlation for a periodic signal is largest when the translation is an integer multiple of the period. However, since the period of the signal is in general not commensurate with the step size $\delta$, we fit the following function around the maximum of $\rho_i$ 
\begin{align}
    \rho_i(\tau\delta) = \rho_i^\t{max}+b\times \t{abs}(\tau_i - \tau_i^\t{max}),
\end{align}
where $\tau$ can take non-integer values and $\tau_i^\t{max}$ corresponds to the period of the signal. The fit parameter $\rho_i^\t{max}$ could be identified with the periodicity of the measurement signal. However, to be more precise, we choose to do another measurement where the sweep range is shifted from the original measurement by $\tau_i^\t{max}\delta$, giving $R_l(f^\prime_i+\tau_i^\t{max}\delta)$. The correlation between this signal and the original one is then computed and denoted as $P$, which is the objective function used in the optimization.  

\section{Experimental implementation}\label{sec:Implementation}
\begin{figure}
    \centering
    \includegraphics[width=4.5in]{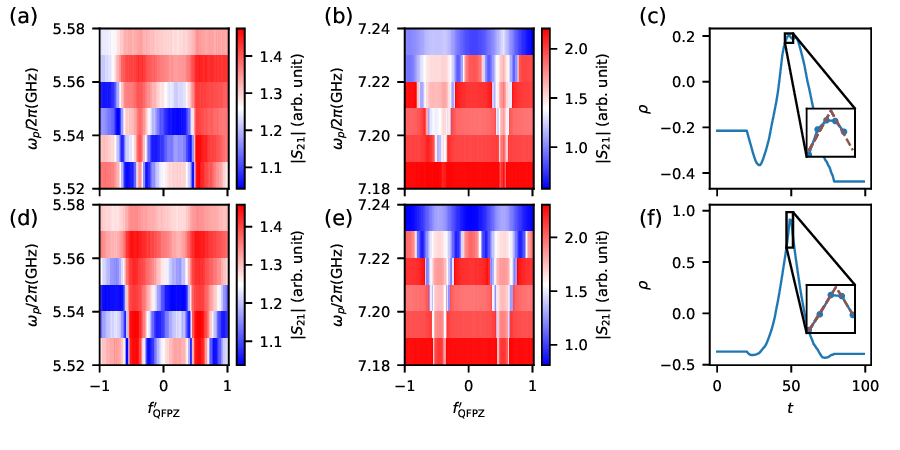}
    \caption{(Color online) Transmission versus probing frequency $\omega_p$ and trial flux coordinate $f_\t{QFPZ}^\prime$, through the fixed-frequency (a,d) and tunable (b,e) resonator, at the first (top) and last (bottom) step in the optimization. The plots on panels (c, f) show the corresponding correlation versus translation, with inset showing the absolute value linear fit around the maxima.}
    \label{fig:TransmissionCorrelation}
\end{figure}

We implement the optimization procedure outlined above on subcircuits of a small prototype coherent quantum annealer. The device consists of two coupled tunable capacitively-shunted flux qubits~\cite{yan_2016_fluxqubitrevisited}, fabricated using a three-stack process in Lincoln Laboratory~\cite{yost_2020_solidstatequbitsintegrated}. Each qubit is coupled to a quantum flux parametron (QFP), which is in turn coupled to a flux-tunable resonator for readout. The QFP acts as an amplifier for the flux signal of the qubit, hence ensuring high-fidelity readout in the qubit flux basis, which is critical for quantum annealing applications~\cite{grover_2020_fastlifetimepreservingreadout}. In addition, each qubit and QFP is inductively coupled to a fixed-frequency resonator to assist crosstalk calibration. A schematic of one qubit unit cell consisting of the qubit, the QFP and the tunable resonator is shown in Fig.~\ref{fig:DeviceFlowChart}(b). The full two-qubit system, including its readout circuits have been calibrated using the iterative translation-based method~\cite{daiCalibrationFluxCrosstalk2021}. The procedure is briefly reviewed in~\ref{app:CISCIQ} and the crosstalk matrix obtained via this method is denoted as the reference crosstalk matrix, $\mb{C}^{\t{ref}}$. 

\subsection{Optimization of a Subcircuit with Three Flux Biases}\label{sec:3LoopOptimization}
For a proof-of-principle demonstration of the periodicity optimization approach, we start with a subcircuit consisting of just the QFP and the tunable resonator. The subcircuit has strong coupling due to the large persistent current of the QFP and the resonator, which makes it very time-consuming to calibrate using the translation-based method. The three flux biases in the subcircuit are denoted as $\t{QFPZ}, \t{QFPX}$ and $\t{TR}$. The optimization starts with the initial crosstalk $\mb{C}^{\t{init}}=\mb{C}^\t{ref}$. This allows setting the qubits and couplers outside the subcircuit in a flux bias such that they are decoupled from the measured subcircuit. Using the reference crosstalk also allows systematic investigation of the performance of the optimization relative to particular initial conditions and bounds on the trial compensation parameters. We have also demonstrated the optimization starting with $\mb{C}^\t{init}$ given by a single iteration of the translation-based method, which is discussed in \ref{app:SingleIteration}.

For the readout channel, we choose to measure transmission through the feedline that is coupled to the resonators. Both the fixed-frequency and tunable resonators are measured, each at six different readout frequencies. The readout frequencies are chosen to be around the bare resonator frequencies and the step size is around their resonance linewidth. The flux bias sweep range is chosen to cover about two periods and the step size is about 20~m$\Phi_0$. The other flux biases not being swept are set to values which avoid the flux-insensitive bias points of the QFP and tunable resonator. This is needed to avoid the tunable resonator and the QFP coincidentally being in flux-insensitive spots, which would make the measurement signal insensitive to crosstalk. Such settings can be achieved without accurate initial estimates of the crosstalk or flux offsets.

As examples for the measurement and analysis at a single step in the optimization, we show the transmission measurement results at the start and the end of the optimization for the QFPZ control periodicity in figure~\ref{fig:TransmissionCorrelation}(a,b,d,e). It is clear that the measurement signal is more periodic after the optimization. This is also reflected in the maximum correlations with respect to translations of the signal, which are shown in figure~\ref{fig:TransmissionCorrelation}(c,f).

\begin{figure*}
    \centering
    \includegraphics[width=4.5in]{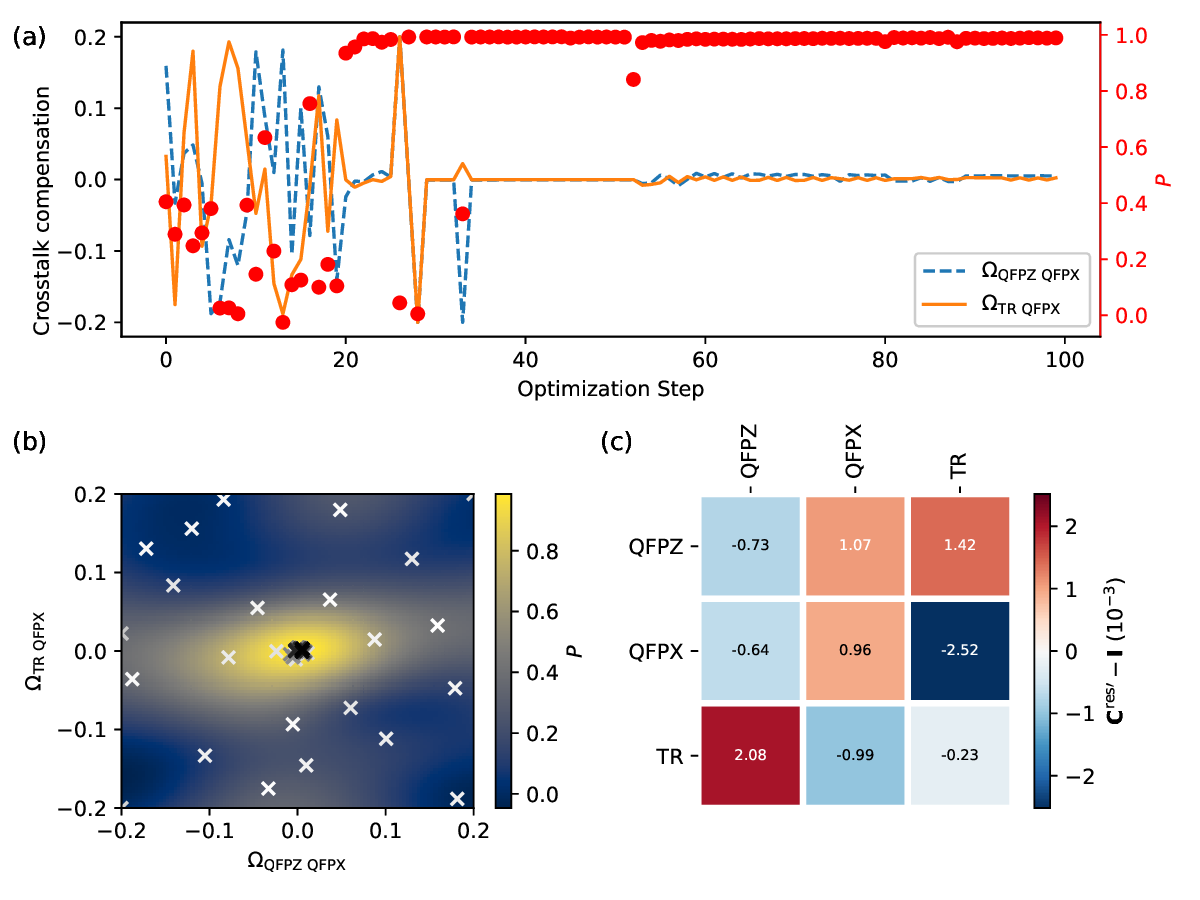}
    \caption{(Color online) (a) Trial compensation parameters (left axis), $\Omega_{\t{QFPZ}~\t{QFPX}}$ (dashed line), $\Omega_{\t{QFPZ}~\t{QFPX}}$ (solid line) and periodicity $P$ (red dots, right axis) versus optimization step. (b) Gaussian process model of periodicity versus the compensation parameters. The cross markers correspond to parameters sampled by the optimizer and the gray scale of the markers indicates the sequence at which they are sampled, with darker color markers being sampled later.(c) Difference between the estimated residual crosstalk matrix $\mb{C}^{\t{res}\prime}$ and the identity matrix.}
    \label{fig:Optimization}
\end{figure*}

We use primarily an optimization algorithm based on Bayesian optimization~\cite{brochuTutorialBayesianOptimization2010}, which is a global optimizer suited for black-box optimization with objective functions which are expensive to evaluate. The algorithm uses a Gaussian process to approximate the objective function, which is called the prior. At each step, the optimizer samples the distribution at a new point in the parameter space, which is probabilistically chosen according to the prior to improve upon the existing samples while minimizing the uncertainties of the prior~\cite{buitinck2013api}. The Gaussian process is then updated according to the Bayesian inference rule, and is used as the prior for the next iteration. The compensation parameters $\Omega_{ji}$'s are bounded to within $[-0.2, 0.2]$, and the optimization is initialized with evaluations at 20 random points in the parameter space. The bounds correspond to typical levels of crosstalk in large-scale devices~\cite{daiCalibrationFluxCrosstalk2021}. We defer to Sec.~\ref{sec:Landscape} for a discussion of how the bounds and initial conditions could affect the optimization. In figure~\ref{fig:Optimization}(a), the trial compensation parameters and the periodicity is plotted versus the optimization step. It can be seen that the optimum parameters have been found after about 40 iterations. There is a drop in the periodicity near step 50, even when the compensation parameters remain mostly unchanged. This is due to the presence of hysteresis in the QFP (see ~\ref{app:Hysteresis} for further discussion). In figure~\ref{fig:Optimization}(b), the landscape of the objective function, predicted by the final Gaussian process model is shown, together with markers for the parameter values sampled during the optimization. The minimum is at around $(\Omega_{\t{QFPZ}~\t{QFPX}}, \Omega_{\t{TR}~\t{QFPX}})=(0, 0)$, which is the expected optimum compensation parameter given $\mb{C}^\t{init}=\mb{C}^\t{ref}\approx \mb{C}$ and hence $\mb{C}^\text{res}\approx \mb{I}$.  In figure~\ref{fig:Optimization}(c), we show the difference between elements of $\mb{C}^{\t{res}\prime}$ with the identity matrix. The magnitudes of the elements are all below $3\times10^{-3}$, which is about the error of the iterative method~\cite{daiCalibrationFluxCrosstalk2021}. This shows that the crosstalk matrix obtained by the optimization method is comparable to the crosstalk matrix obtained by the iterative calibration method.  We also note that the differences are much smaller than the flux sweep step size, which shows that the method does not require high resolution scans to be accurate. As a result of this, the optimization-based measurements required fewer data as compared to one iteration of the translation-based method. 

Using the same optimizer setting but starting the optimization with estimated crosstalk from one iteration of the translation-based method, the estimated crosstalk obtained converged with similar level of accuracy, as compared to starting the optimization with the reference crosstalk matrix, obtained from multiple iterations of translation-based method. The result is presented in~\ref{app:SingleIteration}.

\subsection{Optimization Landscape}\label{sec:Landscape}
\begin{figure}
    \centering
    \includegraphics[scale=0.8]{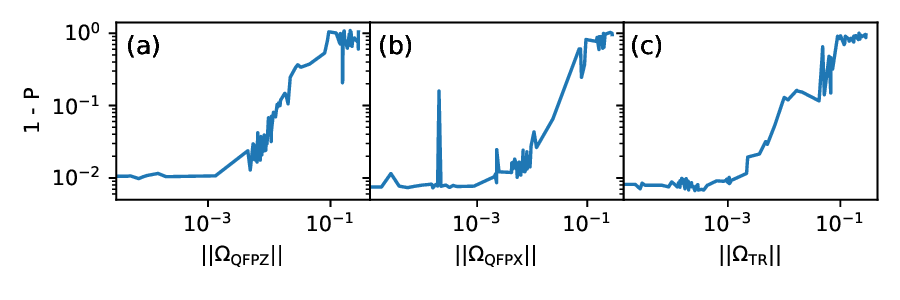}
    \caption{1 minus the periodicity $P$ versus the distance from the trial compensation parameters to the optimized compensation parameters for the three loops, QFPZ (left), QFPX (center), TR (right)}
    \label{fig:OptimizationLandscape}
\end{figure}

After demonstrating that the optimizations converge with high accuracy to the expected compensation parameters, we examine the structure of the optimization problem. First we looked at how periodicity changes as the compensation parameters deviate from the optimized values. We define the distance from the optimized compensation parameters as
\begin{align}
    \lVert \Omega_{i}\rVert =\sum_{j\neq i}(\Omega_{ji}-\Omega_{ji}^\prime)^2
\end{align}
and plot the periodicities measured during the optimization versus $\lVert \Omega_{i}\rVert$ in figure~\ref{fig:OptimizationLandscape}. It can be seen that for all of the loops measured, when $\lVert \Omega_{i}\rVert\lesssim 0.001$, the periodicity function plateaus at about $0.99$, This suggests given the current set of readout channels, the optimized compensation parameters would allow us to control each bias coordinate to $1$~m$\Phi_0$ accuracy over one flux quantum range. The sharp peak for the QFPX loop is likely due to hysteresis of the QFP, which can be avoided by choosing a different set of independent flux control coordinates (see~\ref{app:Hysteresis}). When $\lVert \Omega_{i}\rVert \gtrsim 0.1$, the periodicity $P\approx 0$. This means that the sampled trial compensation parameters are only informative when they satisfy $\lVert \Omega_{i}\rVert \lesssim 0.1$. Hence, the optimization method is likely only efficient when initial crosstalk is known to within $10\%$ accuracy, relative to the diagonal coupling elements. Various sources of estimation could provide such accuracy, such as one single iteration of the translation-based calibration method~\cite{daiCalibrationFluxCrosstalk2021}, measurement on different copies of the same device, or potentially careful electromagnetic simulation of the device. 

\begin{figure}
    \centering
    \includegraphics[scale=0.8]{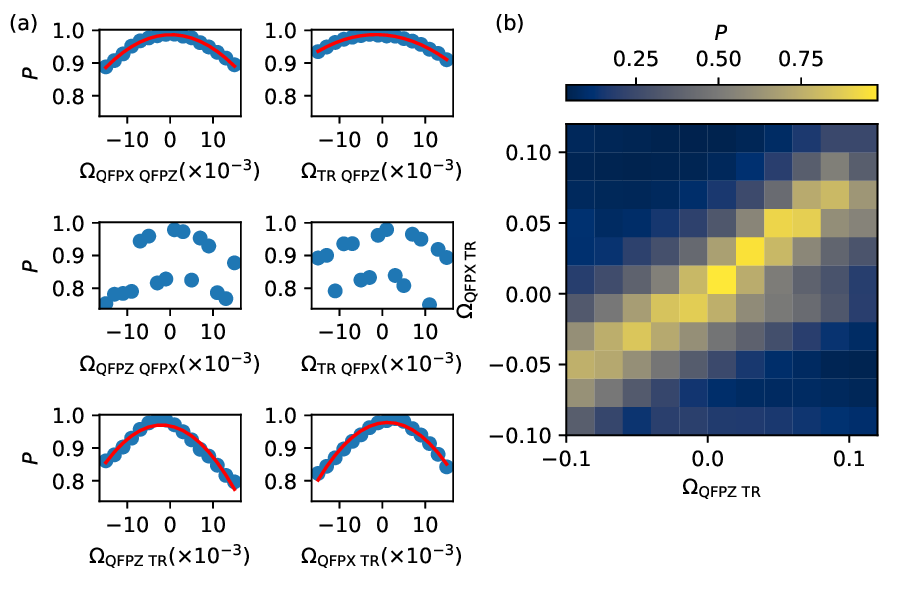}
    \caption{(Color online) (a) Measured (blue dots) periodicity versus deviation of the compensation parameters from the optimized value for each of the six off-diagonal parameters in the $3\times3$ matrix, with a quadratic fit (red curve) on top. (b) Measured periodicity along the $f^\prime_\t{TR}$ bias versus deviation of the compensation parameters $\Omega_{\t{QFPZ}, \t{TR}}, \Omega_{\t{QFPX}, \t{TR}}$ from their optimized values.}
    \label{fig:Landscape}
\end{figure}

We further characterize the landscape of the periodicity function by directly measuring it. This is done by first updating the initial crosstalk matrix with the optimized parameters, according to
\begin{align}
    \mb{C}^\t{init}\xrightarrow{}\mb{C}^{\t{res},\prime}\mb{C}^\t{init},
\end{align}
and then doing measurement in the updated $f^\t{init}$ coordinates. For each loop, the periodicity is measured sweeping one trial compensation parameter, while keeping the other at zero. This measured periodicities are plotted in figure~\ref{fig:Landscape}(a). It can be seen that the periodicity is mostly a smooth function of the compensation parameters with a single maximum. There are two features outstanding. First, the periodicity relative to the QFPX loop has rugged landscape. This is likely due to the QFP becoming hysteric and not responding to flux bias variations fast enough compared to the experiment time. The hysteresis is caused by the discontinuous change in the ground state wavefunction of the QFP at the flux bias symmetry points. This can be systematically avoided by choosing a different set of linearly independent flux bias coordinates, along which the ground state wavefunction changes smoothly (see \ref{app:Hysteresis} for more detailed discussion). Second, the periodicity maxima for the compensations to TR loop are slightly deviated from zero. The reason for this still requires further investigation. One possibility could be that the periodicity function, under the measurement setting used, is sensitive to drifts in flux offsets, which could occur between the optimization measurement and the landscape measurement. The offsets therefore need to be kept track of in future implementations of the optimization, otherwise the accuracy of the crosstalk calibration based on periodicity optimization could be limited. The periodicity along the TR flux bias is also measured sweeping a two-dimensional grid of values for the trial compensations $\Omega_{\t{QFPZ}, \t{TR}}, \Omega_{\t{QFPX}, \t{TR}}$, over the range of $[-0.1, 0.1]$ . The result is plotted in figure~\ref{fig:Landscape}(b). It confirms that the periodicity is a smooth function over the entire range, and has a single maximum at around (0, 0). Such characteristics of the objective function means the optimization problem is likely convex in general. This opens the possibilities of using optimization algorithms that approximates and make use of the local gradients~\cite{spall_1992_multivariatestochasticapproximation,sutskeverImportanceInitializationMomentum2013,lengRobustEfficientAlgorithms2019}. We successfully implement one such optimization method, called simultaneous perturbation stochastic approximation (SPSA)~\cite{spall_1992_multivariatestochasticapproximation} and the result is discussed in~\ref{app:SPSA}.

\subsection{Robustness of the periodicity metric to different measurement ranges}
It is desirable that the periodicity optimization method is robust for a broad range of measurement settings, in particular when limited measurement data is available. In this section, we evaluate the effectiveness of the periodicity metric if the transmission measurement is done with a smaller range of flux bias or fewer frequency values than the measurements presented above. 

\begin{figure}
    \centering
    \includegraphics[width=\linewidth]{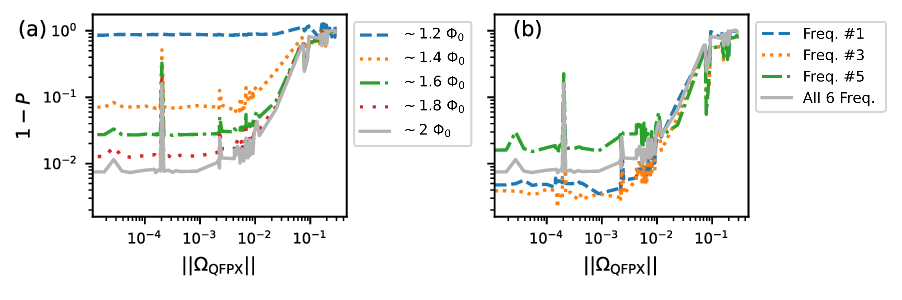}
    \caption{(Color online) (a) 1 minus the periodicity, analyzed with data truncated to a smaller bias range, versus the distance from the trial compensation parameters to the optimized compensation parameters. (b) 1 minus the periodicity, analyzed with data at a chosen frequency for each resonator, versus the distance from the trial compensation parameters to the optimized compensation parameters. In both panels, the control flux coordinate to be optimized is QFPX, and the grey line corresponds to the analysis using the full data (exactly the same as figure~\ref{fig:Landscape}(b). The spikes near $|\Omega_{\rm{QFPX}}|<0.001$ are due to the hysteretic behavior of the QFP. Repeating the analysis for the other two flux coordinates QFPZ and TR yields similar plots. }
    \label{fig:RobustnessBiasFreqRange}
\end{figure}

To obtain the data with smaller flux bias ranges, we take the measurement done at every step of the optimization in section~\ref{sec:3LoopOptimization}, which has a flux bias range of $\sim2~\Phi_0$, truncate the data to a specified smaller range, and then perform the periodicity analysis on the truncated data. Similar to figure~\ref{fig:OptimizationLandscape}, we plot in figure~\ref{fig:RobustnessBiasFreqRange}(a) the periodicity versus the distance from the trial compensation parameters to the optimum compensation parameters, for data truncated to approximately $[1.2, 1.4, 1.6, 1.8]~\Phi_0$ flux bias range respectively. It can be seen that with decreasing flux bias ranges, the maximum periodicity $P$ reached is lower. This is because the overall contrast between the signal at a particular trial flux value $R_l(f_{i,s}^\prime)$ and the average signal $\overline{R_l}$ for a given readout channel $l$ is decreased, reducing the value of the correlation function as given by Equation~\ref{eq:Correlation}. Specifically, for bias range smaller or equal to $\sim 1.4~\Phi_0$, the periodicity metric could not distinguish changes in the compensation parameters on the same order of accuracy as with the full data. However, as long as the bias range is larger than or equal to $\sim 1.6~\Phi_0$, the cost landscape remains qualitatively the same as the full data, with the maxima at $|\Omega_{\rm{QFPX}}|\lesssim 0.003$. The analysis shows that the periodicity maximization does not require multiple periods of bias range, which allows good flexibility in designing the circuit. 

To evaluate the effectiveness of the periodicity metric with fewer frequency values, we again take the data presented in Sec.~\ref{sec:3LoopOptimization} and truncate it to a specified smaller range. It is found that reducing the number of frequency values measured to one per resonator would not significantly alter the cost landscape in the majority of cases. In figure~\ref{fig:RobustnessBiasFreqRange}(b) we show the periodicity versus the distance between the trial compensation parameters to the optimum compensation parameters, evaluated using only one particular frequency for each resonator measurement. It can be seen that when the first frequency value is used, the periodicity achieved is lower, and the landscape becomes noisy near $\Omega_{\rm{QFPX}}\lesssim 0.01$. In contrast, when the third or the fifth frequency value is used, a higher periodicity could be achieved, at a distance $|\Omega_{\rm{QFPX}}|\lesssim 0.003$. \footnote{Using the fourth frequency value yields similar results to using the first, using the zeroth or the second yield similar results to using the third or fifth. They are not shown in the figure for visual clarity.} The periodicity analyzed using the full data corresponds to an "averaged" value of the periodicity obtained using different frequencies, in line with the definition of the correlation function given in equation~\ref{eq:Correlation}. This suggests that the periodicity maximization is not particularly sensitive to the readout frequencies chosen for the measurement, and likely does not require prior fine-tuning. 

\subsection{Optimization of a Subcircuit with Five Flux Biases}\label{sec:5LoopOptimization}
\begin{figure}
    \centering
    \includegraphics[]{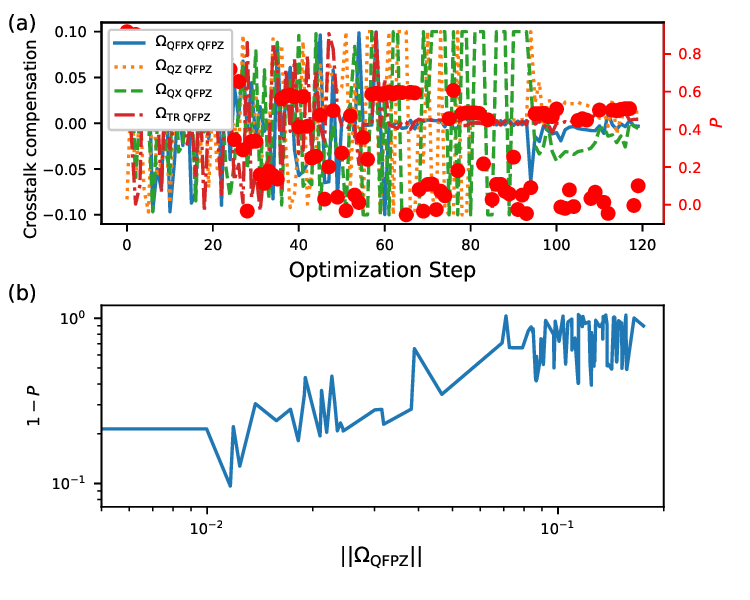}
    \caption{(Color online) Optimization including the qubit loops. (a) Trial compensation parameters (left axis) from QZ, QX, QFX and TR to the QFPZ loop and periodicity $P$ (right axis) versus optimization steps. (b) 1 minus the periodicity $P$ versus the distance from the trial compensation parameters to the optimized compensation parameters for the QFPZ loop. }
    \label{fig:5LoopOptimization}
\end{figure}

To understand the feasibility of the periodicity optimization on larger devices, we implement the procedure incorporating the qubit that is directly coupled to the QFP. The qubit loops are denoted as QZ and QX. In figure~\ref{fig:5LoopOptimization}(a) we show the four compensation parameters and the periodicities with respect to QFPZ loop versus the optimization steps. It is noted that an increased number of initial evaluations, 50, is required for the optimization algorithm to approach the reference compensation parameters. The optimizations for other loops in the system did not approach the reference compensation parameters with the same optimizer hyperparameters. One possible explanation for the relative success of the QFPZ loop periodicity maximization, compared to the other loops, is that the QFPZ loop is special, both because of its large persistent current and it being directly coupled to most other loops in the subcircuit (except QX). The effectiveness of the optimization in larger circuits can potentially be improved by exploring different readout channels and optimization algorithms, which we didn't pursue in this proof-of-principle work.

\section{Discussion and conclusion}\label{sec:Conclusion}
There are a few points worth noting when comparing the iterative translation-based approach and the optimization-based approach. In terms of the number of measurement done at different bias settings, the optimization-based method requires about 3 times less measurements than the translation-based approach to achieve comparable accuracy. The majority of the saving comes from the fact that in the translation-based method, the QFP calibration requires a few high-density 2D scans of its X and Z loop control, while the optimization-based method only requires 1D scans. The translation-based approach cannot use 1D scans to calibrate the crosstalk between the QFP X and Z loop, as the response of the circuit to one of them is highly dependent on the other. This could be considered as an extreme case of two strongly coupled loops (though technically they belong to the same circuit element). Therefore, we expect the optimization-based method could reduce the number of measurements required for calibrating strongly-interacting circuits or multi-loop components, such as the Josephson phase-slip qubit~\cite{kermanSuperconductingQubitCircuit2019}. 

On the other hand, we note that the translation-based approach is more robust than the optimization-based method. The latter requires prior knowledge of the crosstalk matrix to bound the parameter space of the optimization. When the number of optimization parameters increases, the size of the parameter space over which the periodicity metric is insensitive to changes in trial parameters. This leads to an increase in the number of initial evaluations required for the Bayesian algorithm to learn the cost landscape, as we see in section~\ref{sec:5LoopOptimization}. Therefore, we expect that more initial knowledge of the crosstalk matrix is required in order to implement the current Bayesian optimization-based periodicity maximization to larger devices. 

In summary, we introduced a flux crosstalk calibration method based on the fundamental periodicity of superconducting circuits, without relying on any underlying circuit model. The method is successfully demonstrated on a coupled QFP-tunable-resonator system, with an accuracy that is comparable to the previously developed iterative translation-based calibration method. Although the current implementation of the optimization approach is limited when used to calibrate devices with larger number of loops, it can already be utilized as a subroutine for calibrating parts of a larger system. Such hybrid calibration strategy is particularly useful for strongly interacting systems such as the quantum annealing circuits investigated here, where translation-based approach alone requires larger number of iterations and high-resolution measurements.

The landscape measurement shows that the problem is nearly convex within some bounds on the optimization parameters. This points to exploring other optimization algorithms, such as momentum based optimizations~\cite{sutskeverImportanceInitializationMomentum2013,lengRobustEfficientAlgorithms2019} to speed up the convergence, which is crucial for extending the optimization-based calibration to larger devices. Another attractive future direction could be adaptive measurements, where different experiment parameters can be used to give different optimization landscapes. For example, an optimization landscape with a broad maximum could afford large tolerance to the initial guess of the crosstalk matrix, while optimization landscape with a narrow maximum could lead to higher accuracy for the optimized results.

\ack
We thank the members of the Quantum Enhanced Optimization (QEO)/Quantum Annealing Feasibility Study (QAFS) collaboration for various contributions that impacted this research. In particular, we would like to acknowledge S. Novikov for designing the device measured in this work, A. Martinez for useful discussion on Bayesian optimization, D. M. Tennant for the help with experiment infrastructure, D. Ferguson for discussion on the periodicity properties, J. A. Grover for discussion on the measurement code, A. J. Kerman for help with circuit modelling and K. M. Zick for leading the QEO/QAFS experimental effort. We gratefully acknowledge the MIT Lincoln Laboratory design, fabrication, packaging, and testing personnel for valuable technical assistance. The research is based upon work supported by the Office of the Director of National Intelligence (ODNI), Intelligence Advanced Research Projects Activity (IARPA) and the Defense Advanced Research Projects Agency (DARPA), via the U.S. Army Research Office contract W911NF-17-C-0050. The views and conclusions contained herein are those of the authors and should not be interpreted as necessarily representing the official policies or endorsements, either expressed or implied, of the ODNI, IARPA, DARPA, or the U.S. Government. The U.S. Government is authorized to reproduce and distribute reprints for Governmental purposes notwithstanding any copyright annotation thereon.

\begin{appendix}
\section{CISCIQ iteration}\label{app:CISCIQ}
In this section we briefly review the procedure of the iterative crosstalk calibration method introduced in Ref.~\cite{daiCalibrationFluxCrosstalk2021}, named CISCIQ (crosstalk into SQUIDs, crosstalk into Qubits). In the first iteration, measurements are done by sweeping each control voltage and measuring the response of each loop, with the assumption that the effect of circuit interactions can be averaged out and the measured response can be completely attributed to the voltage to flux coupling coefficient $C_{ij}$. The first iteration hence gives the first estimate of the crosstalk matrix and flux offsets, denoted as $\mb{C}^{(1)\prime}$ and $\mb{f}_0^{(1)\prime}$. The second iteration performs the same measurement as the first iteration, but sweeping the estimated flux coordinates
\begin{align}
    \mathbf{f}^{(1)}=\mathbf{C}^{(1) \prime} \mathbf{V}+\mathbf{f}_{0}^{(1)\prime}.
\end{align}
After $n$ iterations, the final estimated matrix is
\begin{align}
    \mathbf{C}^{\t{ref}}=\mathbf{C}^{(n) \prime} \mathbf{C}^{(n-1) \prime} \ldots \mathbf{C}^{(1) \prime},
\end{align}
where the superscript denotes that this is the reference crosstalk value we use to compare with the crosstalk matrix obtained by periodicity optimization. It is expected that each new iteration measures a smaller residual crosstalk and flux offsets, so that the estimates converges towards the true values. 

\begin{figure}
    \centering
    \includegraphics[width=4.5in]{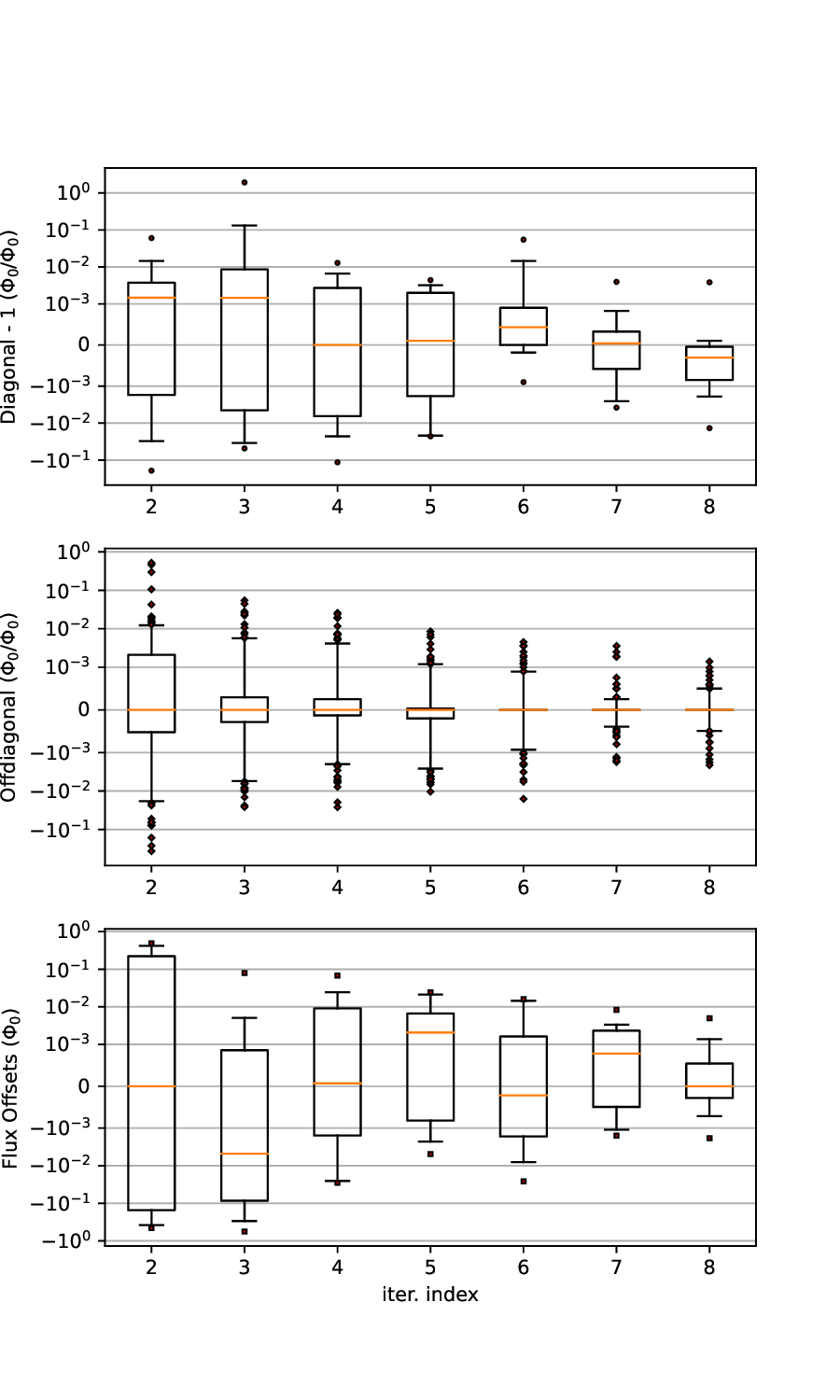}
    \caption{For the full two-qubit device, statistical box plots of the diagonal (top), off-diagonal (middle) coupling coefficients in $\mb{C}^{(n)\prime}$, and the flux offsets (bottom) in $\mb{f}^{(n)\prime}_0$ versus the iteration number. The orange bar is the median, the black box corresponds to the lower and upper quartiles, the segments contain the 5th to 95th percentiles of the data and the dots are outliers.}
    \label{fig:CISCIQConvergence}
\end{figure}

\begin{figure}
    \centering
    \includegraphics[width=4.5in]{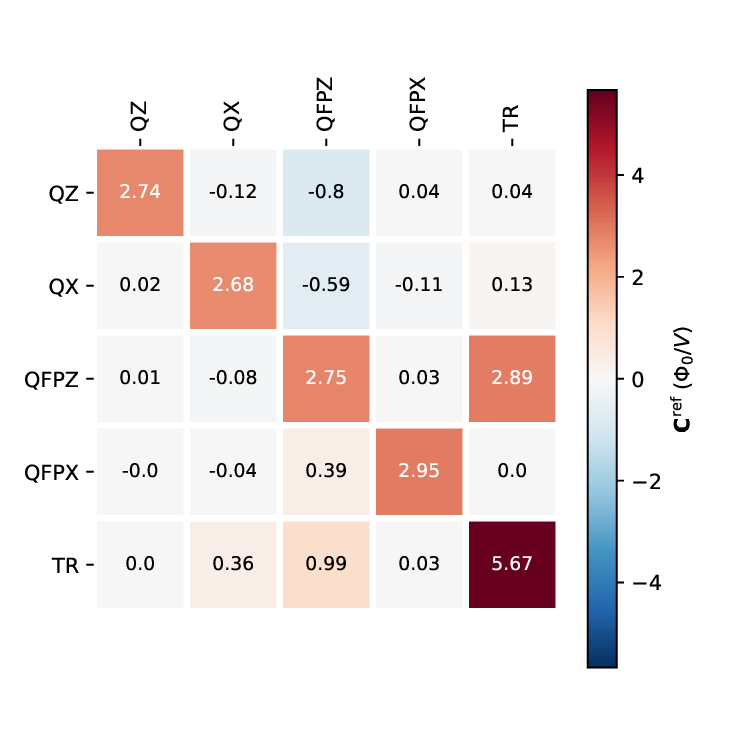}
    \caption{Final crosstalk matrix $\mathbf{C}^{\t{ref}}$ obtained from the iterative procedure for the qubit, QFP and tunable resonator subcircuit, on which the periodicity optimization is applied. }
    \label{fig:CISCIQResults}
\end{figure}

The iterative approach is applied to the two-qubit circuit described in the main text. In figure~\ref{fig:CISCIQConvergence} we show how the measured crosstalk and flux offsets converge towards identity and zeros. In figure~\ref{fig:CISCIQResults} we show the final crosstalk matrix from the iterative procedure for the qubit, QFP and tunable resonator, on which the periodicity optimization approach is implemented. 

\section{Optimization initialized with single iteration of translation-based calibration}\label{app:SingleIteration}
\begin{figure}
    \centering
    \includegraphics{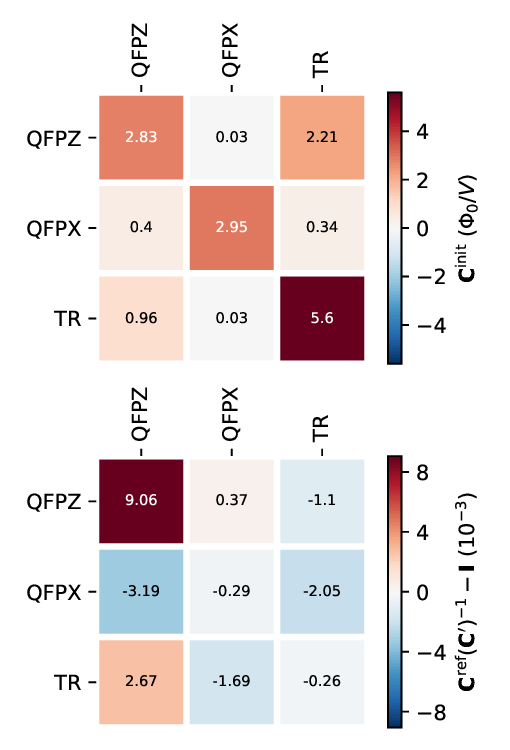}
    \caption{(Color online) Top: $\mb{C}^\t{init}$ as given by the first iteration of the translation-based calibration method. Bottom: The deviation between the optimized crosstalk matrix $\mb{C}^\prime$ and the reference crosstalk matrix $\mb{C}^\text{ref}$, defined as $\mb{C}^\text{ref}(\mb{C}^\prime)^{-1}-\mb{I}$. The optimized crosstalk matrix $\mb{C}^\prime=\mb{C}^{\t{res}\prime}\mb{C}^\t{init}$ is obtained by starting the optimization with the matrix given in the top panel.}
    \label{fig:3LoopCISCIQ0}
\end{figure}

In this section we describe the results obtained by performing the periodicity optimization, starting from the estimated crosstalk of one iteration of the translation-based approach. In figure~\ref{fig:3LoopCISCIQ0}(a) we show the estimated crosstalk matrix obtained by a single iteration. This can be compared with the reference matrix elements plotted in figure~\ref{fig:CISCIQResults}. It can be seen that after a single iteration, the estimated crosstalk still deviates from the reference values, by as large as $\sim 10\%$.

In figure~\ref{fig:3LoopCISCIQ0}(b), we plot the deviation between the reference crosstalk matrix and the estimated crosstalk matrix after the optimization. Most of the deviation is about or less than $3\times10^{-3}$. This is comparable to the accuracy of the results discussed in the main text, which starts the optimization directly from $\mb{C}^{\t{ref}}$. The only exception is the QFPZ diagonal element, which corresponds to its period. This is probably due to the hysteresis of the QFP, which can be resolved by repeating the QFPZ periodicity measurement at a different QFPX biasing point. 

\section{Evidence for hysteresis of QFP and how to resolve it}\label{app:Hysteresis}
\begin{figure}
    \centering
    \includegraphics{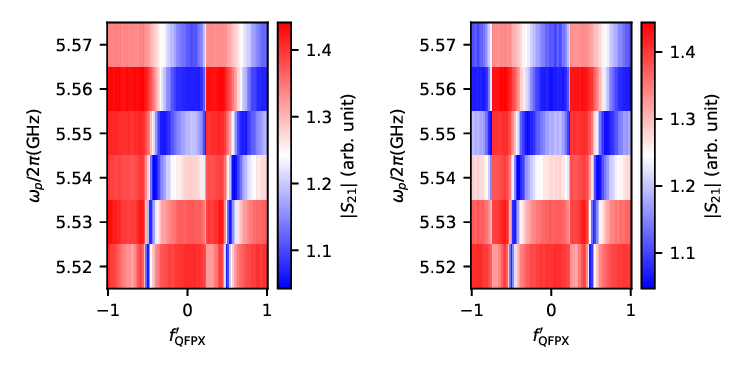}
    \caption{Transmission versus the probing frequency and the QFPX bias, at two different values of the compensation parameter $\Omega_{\t{QFPZ}~\t{QFPX}}=-0.001$ (left) and $0.001$ (right).}
    \label{fig:Hysteresis}
\end{figure}
It is found that the optimization landscape for the periodicity of QFPX loop is often not well-behaved. 
As mentioned in the main text, this is attributed to the hysteresis of the QFP. In this section we briefly discuss the evidence of the hysteresis in the data and the solution to this problem based on the double-well potential of the QFP.

In figure~\ref{fig:Hysteresis}(a) we show the transmission versus the probing frequency and the QFPX bias at two different values of the compensation parameter $\Omega_{\t{QFPZ}~\t{QFPX}}=\pm0.001$, during the landscape measurement discussed in Sec.~\ref{sec:Landscape}. It can be seen that when $\Omega_{\t{QFPZ}~\t{QFPX}}=0.001$, there are three periodically separated resonance traces while when $\Omega_{\t{QFPZ}~\t{QFPX}}=-0.001$, the resonance trace at around $f^\prime_\t{QFPX}\approx -1$ is missing. This suggests that QFP is not responding to the flux bias variations within the experiment time. 

The QFP, can be approximately described by a two-level Hamiltonian in the persistent current basis, given by
\begin{align}
    H=-I_{p}(f_{Z}-f_{Z,\t{sym}}) \Phi_{0} \sigma_{z}-\frac{\Delta}{2} \sigma_{x},
\end{align}
where $I_p$ is the persistent current in the Z loop, $\Delta$ is the tunneling amplitude controlled by $f_X$, and we use Z, X instead of QFPZ, QFPX to simplify notation. The symmetry point of the QFP $f_{Z,\t{sym}}$, just like the tunable CSFQ and the rf-SQUID qubit, is given by $f_{Z,\t{sym}}=1/2+1/2 f_X$, assuming junction asymmetry is negligible. This means when the X flux is being swept, both the tunneling and the biasing between the two persistent current states are changing. Due to the large persistent current of the QFP ($\sim 1\mu$A), there is a region in flux bias where the tunneling is small and the QFP could not tunnel to the persistent current state with lower energy. To solve this problem, we can replace the Z and X biases with 
\begin{align}
    \tilde{f}_\t{Z}&=f_\t{Z}+\frac{1}{2}f_\t{X},\\
    \tilde{f}_\t{X}&=f_\t{X}.
\end{align}
By keeping $\tilde{f}_\t{Z}$ approximately fixed when sweeping $\tilde{f}_\t{X}$, one can avoid switching the sign of the bias between the two persistent current states, and hence avoiding the need for tunneling to occur for the QFP to respond to changes in flux biases. Using the new flux bias coordinates, the periodicity along $\tilde{f}_\t{X}$ increases to $2$ (keeping $\tilde{f}_\t{Z}$ fixed). 

\section{Optimization with SPSA}\label{app:SPSA}
\begin{figure}
    \centering
    \includegraphics{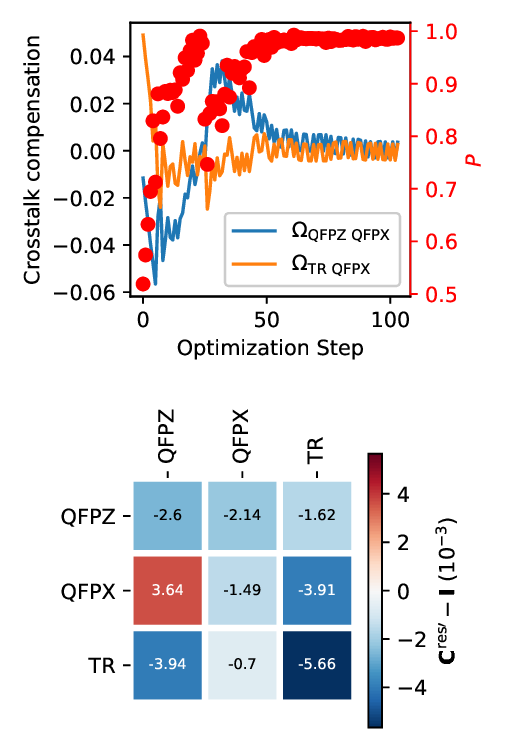}
    \caption{(Color online) Top: Trial compensation parameters (left axis), $\Omega_{\t{QFPZ}~\t{QFPX}}$, $\Omega_{\t{TR}~\t{QFPX}}$ and periodicity $P$ (right axis) versus SPSA optimization step. Bottom: Difference between the estimated residual crosstalk matrix $\mb{C}^{\t{res}\prime}$, obtained using SPSA optimization, and the identity matrix.}
    \label{fig:SPSA}
\end{figure}

In this section we discuss the optimization results using an alternative optimizer called the Simultaneous Perturbation Stochastic Approximation (SPSA)~\cite{spall_1992_multivariatestochasticapproximation}. This algorithm approximates the gradient of the objective function by measuring the finite difference due to a perturbation vector along a random direction in the parameter space, and performs gradient descent. We start the SPSA optimization with $\mb{C}^\t{ref}$ and the initial point is chosen uniformly randomly in the range $[-0.1, 0.1]$. In figure~\ref{fig:SPSA}(a) a typical optimization process is shown, plotting the compensation parameters to QFPX and the periodicity versus the optimization step. The optimization converges after about 60 iterations and oscillates afterwards. The optimized compensation parameters are used to compute the estimated $\mb{C}^{\t{res}\prime}$ and its difference from the identity matrix is shown in figure~\ref{fig:SPSA}(b). The difference is about twice as large as compared to the results obtained using Bayesian optimization. We expect the results to improve by using better hyper-parameters for the optimization, such as the magnitude of the perturbation, which would likely remove the parameter oscillations near the end of the optimization.

\end{appendix}

\newpage
\bibliography{references.bib}
\bibliographystyle{iopart-num.bst}

\end{document}